\documentclass{ws-mplb}
\usepackage{latexsym}
\usepackage{revsymb}
\usepackage{amsfonts}
\usepackage{amsmath}
\usepackage{amssymb}
\usepackage{bm}

\newcommand{\vev}[1]{\langle #1\rangle}
\newcommand{\bra}[1]{\langle #1 |}
\newcommand{\ket}[1]{| #1 \rangle}
\newcommand{\braket}[2]{\langle #1 | #2\rangle}

\newcommand{\0}{\underline{0}}
\newcommand{\1}{\underline{1}}

\newcommand{\uk}{\underline{k}}
\newcommand{\Fq}{\mathbb{F}_q}
\newcommand{\Fone}{\mathbb{F}_{1}}
\newcommand{\Fun}{\mathbb{F}_\mathrm{un}}

\def\draftnote{\today\quad\currenttime\hfill Report Number: IPMU13-0235}
\begin{document}

\title{
QUANTUM $\bm{\Fun}$ :\\ 
THE $q=1$ LIMIT OF GALOIS FIELD QUANTUM MECHANICS,
PROJECTIVE GEOMETRY \& THE FIELD WITH ONE ELEMENT
}

\author{\footnotesize LAY NAM CHANG}
\address{Department of Physics, Virginia Tech, Blacksburg, VA 24061, USA\\
laynam@vt.edu}

\author{\footnotesize ZACHARY LEWIS}
\address{Department of Physics, Virginia Tech, Blacksburg, VA 24061, USA\\
zlewis@vt.edu}

\author{\footnotesize DJORDJE MINIC}
\address{Department of Physics, Virginia Tech, Blacksburg, VA 24061, USA\\
dminic@vt.edu}

\author{\footnotesize TATSU TAKEUCHI}
\address{Department of Physics, Virginia Tech, Blacksburg, VA 24061, USA
, and\\
Kavli Institute for the Physics and Mathematics of the Universe (WPI),\\The University of Tokyo, Kashiwa-shi, Chiba-ken 277-8583, Japan\\
takeuchi@vt.edu}

\maketitle

\begin{abstract}
We argue that the $q=1$ limit of Galois Field Quantum Mechanics,
which was constructed on a vector space over the Galois Field $\mathbb{F}_q=GF(q)$,  
corresponds to its `classical limit,'
where superposition of states is disallowed.
The limit preserves the projective geometry nature of the state space,
and can be understood as being constructed on an appropriately defined analogue of 
a `vector' space over the 
`field with one element' $\Fone$.
\end{abstract}


\ccode{03.65.Aa,03.65.Ta,03.65.Ud}

\section{Introduction - the Classical Limit of Quantum Mechanics}

How deterministic classical mechanics (CM) emerges from probabilistic quantum mechanics (QM)
is a conundrum which is yet to be resolved.
As first noticed by Dirac\cite{Dirac:1930},
the Poisson brackets of CM can be considered the $\hbar\rightarrow 0$ 
limit of commutators in QM,
and the Heisenberg equation goes over to Hamilton's equation in that limit:
\begin{equation}
\dfrac{d\hat{A}}{dt} \;=\; \dfrac{1}{i\hbar}[\,\hat{A},\,\hat{H}\,]
\qquad\xrightarrow{\hbar\rightarrow 0}\qquad
\dfrac{dA}{dt} \;=\; \{\,A,\,H\,\}\;.
\end{equation}
Despite this formal correspondence, however,
it is far from clear how the full classical theory would be recovered from
QM in such a limit.
In particular, what happens to the Hilbert space of QM in the $\hbar\rightarrow 0$ limit is a difficult question.\footnote{%
The singularity of the $\hbar\rightarrow 0$ limit
has been emphasized in an illuminating review by Berry\cite{Berry:2001}.
}
Also, $\hbar$ being a dimensionful quantity, the limit $\hbar\rightarrow 0$ itself is not particularly well-defined.\footnote{%
Given the dimensionfulness of $\hbar$, 
the $\hbar\rightarrow 0$ limit should be understood as
$\hbar\ll S$ where $S$ is the classical action that is used to describe the
quantum formulation as, say, in the path integral approach. 
}

In order to see the CM$\leftrightarrow$QM correspondence more clearly,
several attempts have been made to make CM look more like QM via
the introduction of wave-functions, operators, and probability distributions.
The WKB approximation\cite{Wentzel:1926,Kramers:1926,Brillouin:1926}
to the Schr\"odinger equation has been reinterpreted as the wave-function of
CM by Van Vleck\cite{VanVleck:1928} and subsequently by Schiller,\cite{Schiller:1962a,Schiller:1962b,Schiller:1962c} with observables represented by commuting hermitian operators with continuous eigenvalues.
However, utilizing the rewriting of the Schr\"odinger equation 
by de~Broglie in his pilot-wave theory,\cite{Bacciagaluppi:2013} which was later elaborated
on by Bohm to discuss a hidden variable interpretation of QM,\cite{Bohm:1952a,Bohm:1952b}
Rosen has shown 
that taking the $\hbar\rightarrow 0$ limit of the Schr\"odinger equation does not
necessarily recover the Hamilton-Jacobi equation
for superpositions of states.\cite{Rosen:1964,Rosen:1965,Rosen:1986}

Koopman\cite{Koopman:1931} and von~Neumann\cite{vNeumann1:1932,vNeumann2:1932},
and later Sudarshan\cite{Sudarshan:1976},
developed a complete formulation of CM on a Hilbert space, which was again characterized by commuting hermitian operators as observables.
Recent work on extending the Koopman-von Neumann-Sudarshan formalism with the introduction of path integrals, etc. include Refs.~\refcite{Gozzi:2001he,Mauro:2001rm,Mauro:2003,Gozzi:2003sh,Carta:2005fq}.
It should be emphasized that the Koopman-von Neumann-Sudarshan theory is not the $\hbar\rightarrow 0$ limit of QM.
Indeed, the operators that correspond to position and momentum in their formalism 
commute with each other without the taking of any limit, 
and respectively have canonically conjugate operators with which they do not
commute but are deemed unobservable.
The superposition of states also correspond to ensembles of classical states, 
and not macroscopic Schr\"odinger-cat like states.
Thus, though the Koopman-von Neumann-Sudarshan theory is a formulation of CM on a Hilbert space,
it does not (yet) provide much insight on how CM can emerge from QM, or vice versa.

Given the apparent absence of macroscopic Schr\"odinger-cat like states,
they must somehow vanish in the classical limit.
Several approaches have been used to address this problem, the most prominent being
that of `environmental decoherence' reviewed in Refs.~\refcite{Zurek:2003,Zurek:2003zz} by Zurek.
There, the system under observation and its environment
(rest of the universe) are both treated quantum mechanically, and it has been shown that
the interaction between the two suppresses the off-diagonal terms in the density matrix
in a preferred basis. 
The statistical nature of the theory remains, however, and 
leads to the many-worlds interpretation of QM developed from the pioneering work of Everett\cite{Everett:1957hd}.
Other approaches to the system-environment interaction problem
treat the environment classically, and the interaction with the
quantum system leads to classical-quantum `hybrid' theories, the properties of which are still under
investigation.\footnote{%
See Refs.~\refcite{Sudarshan:1976,Sherry:1978ea,Sherry:1979vc,Gautam:1978ws,Peres:2001,Terno:2004ti,Elze:2011hi,Elze:2012fk,Elze:2013una,Jauslin:2011,Salcedo:2012qx,Barcelo:2012ja} and
references therein.}

Thus, despite impressive developments in our understanding of QM during the past century,
the CM/QM divide remains and bridging that gap is still an intense area of investigation.
What seems probable is that recovering CM from QM would require more than
taking the $\hbar\rightarrow 0$ limit of QM in some yet to be discovered way,
or the rewriting of CM or QM to look more like the other in some fashion.
A more generic theory which encompasses both CM and QM may be necessary to understand the
CM$\leftrightarrow$QM correspondence.

As with any difficult problem, finding a toy analog which
simplifies the situation while maintaining the essence of the quandary
is often instructive.
In Refs.~\refcite{Chang:2012eh} and \refcite{Chang:2012gg}
we constructed a toy analog of $N$-level QM on the
vector space $\Fq^N$,
where $\Fq=GF(q)$ is the Galois field of order $q=p^n$,
with $p$ prime and $n\in\mathbb{N}$,
which we dubbed
``Galois Field Quantum Mechanics'' (GQM).\footnote{%
See also Refs.~\refcite{MQT,MQT2,MQT3,MQT4,James:2011,Willcock:2011,Hanson:2011,Hanson:2012,Hanson:2013,Chang:2012we,Takeuchi:2012mra,Chang:2013joa,Ellerman:2013,Ellerman:2014}.
}
GQM was necessarily different from canonical QM in many ways, 
due to the vector space $\Fq^N$ not possessing an inner product.
Consequently, observables were not represented by operators, hermiticity being difficult to
define without an inner product\footnote{%
It is possible to define analogs of hermitian operators using biorthogonal systems.
See Refs.~\refcite{Chang:2012we,Curtright:2005zk,Curtright:2006zv}.},
and without operators there were no commutators, or $\hbar$ for that matter.
However, in it, physical states were still represented by 
elements of a projective geometry, and the theory still predicted
probabilities of the outcomes of a measurement which could not
be mimicked by any hidden variable theory.
Thus GQM, despite being constructed on a discrete and finite vector space $\Fq^N$ without
an inner product, nevertheless captured some of the quantum-ness
of canonical QM.
A natural question to ask then is: does GQM have a `classical' limit
in which this quantum-ness is lost and replaced by classical-ness?
An answer to this could help us understand how canonical QM becomes CM also.

In this paper, we present the observation that even though GQM does not have an $\hbar$
in its formulation,
its `classical' limit can still be defined by taking the limit $q\rightarrow 1$.
In that limit, the Galois Field $\Fq$ can be expected to become $\Fone$, aka $\Fun$,\footnote{%
Denoting $\Fone$ as $\Fun$ is a French-English bilingual pun. See, e.g. Refs.~\refcite{LeBruyn:2009,Connes:2008a}.}
the `field with one element,' 
an exotic and somewhat nebulous mathematical concept first suggested by Tits in 1957.\cite{Tits:1957}
$\Fq$ in turn can be considered the `quantum'-deformation of $\Fone$.\cite{qAnalog}
Though dormant for many decades, the study of $\Fone$, and efforts to actually define what it is,
has intensified since the 1990's\footnote{%
See, for instance, Refs.~\refcite{Manin:1995,Manin:2008,Zhu:2000,Soule:2003,Soule:2009,Kurokawa:2003,Kurokawa:2005,Koyama:2012,Deitmar:2005,Deitmar:2006a,Deitmar:2006b,Deitmar:2008,Borger:2009,Lorscheid:2009,LopezPena:2009,LeBruyn:2009,Connes:2008a,Connes:2008b,Connes:2009a,Connes:2009b,Connes:2010a,Connes:2010b,Lescot:2009,Lescot:2011,Lescot:2012}.} 
under the expectation that it would lead to a proof of the Riemann hypothesis.\footnote{%
According to Kurokawa and Koyama in Ref.~\refcite{Kurokawa:2010},
the proof for the analogue of the Riemann hypothesis on
projective algebraic varieties over $\mathbb{F}_q$
is known (one of the Weil conjectures proved by Deligne in 1974).
The hope is that reinterpreting the integers $\mathbb{Z}$ as
a projective algebraic variety over $\Fone$
will lead to a proof of the Riemann hypothesis utilizing similar techniques.
In a separate popular book\cite{Kurokawa:2013}, Kurokawa and Koyama state that
Mochizuki's recent work on the ABC conjecture\cite{Mochizuki:2012} also uses ideas based on $\Fone$.}\footnote{%
See also Ref.~\refcite{BerryKeating:1998} for a quantum mechanical approach to the
Riemann hypothesis.
}

This paper is organized as follows.
In section~2 we review how GQM is constructed on $\Fq^N$.
We will look at what happens to the theory if we let $q=1$ for the case $N=2$,
and discuss its `classical' properties.
In section~3 we review the notion of $\Fone$, the `field with one element,'
following the treatment of Kurokawa and Koyama,\cite{Kurokawa:2010}
and show that the $q=1$ limit of GQM can be constructed directly 
on the `vector' space $\Fone^N$.
The resulting state space is a projective geometry, just as with GQM and
canonical QM.
The theory on $\Fone^N$ also prohibits the superposition of states,
precisely the property one expects in a `classical' theory.
Discussion on what this teaches us is given in section~4.

\section{Galois Field Quantum Mechanics}

\subsection{The Model}

Since the details of GQM can be found in Refs.~\refcite{Chang:2012eh} and \refcite{Chang:2012gg},
here we only present the basic outline.

Consider the vector space $\Fq^N$.
Vectors in $\Fq^N$ represent states of the model system, while
dual vectors in the dual vector space $\Fq^{N*}$ represent possible
outcomes of measurements.   
The probability of obtaining the outcome represented by the dual-vector $\bra{x}\in\Fq^{N*}$
when a measurement is performed on the state represented by the 
vector $\ket{\psi}\in\Fq^N$ is given by 
\begin{equation}
P(x|\psi) \;=\; \dfrac{\bigl|\braket{x}{\psi}\bigr|^2}{\sum_y \bigl|\braket{y}{\psi}\bigr|^2}\;,
\label{Pdef}
\end{equation}
where the sum in the denominator runs over all the dual vectors $\bra{y}$ 
in a basis of $\Fq^{N*}$ which includes $\bra{x}$.
The choice of basis of $\Fq^{N*}$ corresponds to an observable, each dual vector
in the basis representing a different outcome.

The absolute value function in the above expression
converts elements of $\Fq$ into either zero or one in $\mathbb{R}$:
\begin{equation}
|\,\underline{k}\,|\;=\;
\left\{\begin{array}{ll}
0\quad &\mbox{if $\underline{k}=\0$}\;,\\
1\quad &\mbox{if $\underline{k}\neq\0$}\;.
\end{array}
\right.
\label{abs}
\end{equation}
Here, underlined numbers and symbols represent elements of $\Fq$, to
distinguish them from elements of $\mathbb{R}$ or $\mathbb{C}$.
Since $\Fq\backslash\{\0\}$ is a cyclic multiplicative group, 
this assignment of `absolute values' is the only one consistent with 
the requirement that the map from $\Fq$ to non-negative $\mathbb{R}$ be product preserving,
that is: $|\underline{k}\underline{l}|=|\underline{k}||\underline{l}|$.\footnote{%
The product preserving nature of the absolute value function 
guarantees that the probabilities of product observables on product states
factorize in multi-particle systems.
This property is crucial if we want to have isolated particle states.
}
Since the same absolute value is assigned to all non-zero brackets,
all outcomes $\bra{x}$ for which the bracket with the state $\ket{\psi}$
is non-zero are given equal probabilistic weight.

Note also that the multiplication of $\ket{\psi}$ with a non-zero element of
$\Fq$ will not affect the probability. 
Thus,
vectors that differ by non-zero multiplicative constants are identified as representing the same 
physical state, and the state space is endowed with the finite projective geometry
\cite{Hirschfeld,Hirschfeld2,Arnold,Ball-Weiner}
\begin{equation}
PG(N-1,q) \;=\; (\,\Fq^N\backslash\{\mathbf{\0}\}\,)\,\big/\,(\,\Fq\backslash\{\0\}\,)\;,
\end{equation}
where each `line' going through the origin of $\Fq^N$ is identified as a `point,'
in close analogy to the complex projective geometry $\mathbb{C}P^{N-1}$ 
of canonical $N$-level QM defined on $\mathbb{C}^N$.

\subsection{An Example}

To give a concrete example of our proposal, let us construct a 2-level system,
analogous to spin, on $\Fq^2$ for which the state space is
$PG(1,q)$.
This geometry consists of $q+1$ `points,'
which can be represented by the vectors
\begin{equation}
\ket{\,0\,} = \left[\begin{array}{c} \1 \\ \0 \end{array}\right],\quad
\ket{\,1\,} = \left[\begin{array}{c} \0 \\ \1 \end{array}\right],\quad
\ket{\,r\,} = \left[\begin{array}{l} \underline{a}^{r-1} \\ \1 \end{array}\right],
\label{kets}
\end{equation}
$r=2,3,\cdots,q$, where $\underline{a}$ is the generator of the multiplicative group $\Fq\backslash\{\0\}$
with $\underline{a}^{q-1}=1$.
The number $q+1$ results from the fact that of the $q^2-1$ non-zero vectors in $\Fq^2$, 
every $q-1$ are equivalent,
thus the number of inequivalent vectors is $(q^2-1)/(q-1)=(q+1)$.
Similarly, the $q+1$ inequivalent dual-vectors can be represented as:
\begin{eqnarray}
\bra{\,\overline{0}\,} & = & \bigl[\,\0\;-\!\1\,\bigr]\;,\cr
\bra{\,\overline{1}\,} & = & \bigl[\,\1\;\;\phantom{-}\0\,\bigr]\;,\cr
\bra{\,\overline{r}\,} & = & \bigl[\,\1\;-\!\underline{a}^{r-1}\,\bigr]\;,\qquad r=2,3,\cdots,q\;,
\label{bras}
\end{eqnarray}
where the minus signs are dropped when the characteristic of $\Fq$ is two.\footnote{\label{characteristic}%
The `characteristic' of a field is the smallest non-negative integer $m$
such that $\underbrace{\1+\1+\cdots +\1}_{m\;\mathrm{times}}=\0$,
where $\1$ is the multiplicative unit and $\0$ is the additive unit.
For example, the characteristic of $\Fq$ with $q=p^n$ is the prime $p$.
The characteristics of $\mathbb{Q}$, $\mathbb{R}$, and $\mathbb{C}$ are
defined to be zero.
}
From these definitions, we find:
\begin{eqnarray}
\braket{\bar{r}}{s} 
& =    & \0\quad \mbox{if $r=s$}\;, \cr 
& \neq & \0\quad \mbox{if $r\neq s$}\;,
\end{eqnarray}
and
\begin{equation}
\bigl|\braket{\bar{r}}{s}\bigr|\;=\; 
\begin{cases}
0 & \mbox{if $r=s$}\;, \\
1 & \mbox{if $r\neq s$}\;. 
\end{cases}
\label{braketpq}
\end{equation}
Observables are associated with a choice of basis of $\Fq^{2*}$:
\begin{equation}
A_{rs}\;\equiv\;\{\;\bra{\bar{r}},\;\bra{\bar{s}}\;\}\;,\qquad
r\neq s\;.
\label{Ars}
\end{equation}
We assign the outcome $+1$ to the first dual-vector of the pair, 
and the outcome $-1$ to the second to make these observables
spin-like. This assignment implies $A_{sr}=-A_{rs}$.
The indices $rs$ can be considered as indicating the direction of the `spin,'
and the interchange of the indices as indicating a reversal of this direction.

Applying Eq.~(\ref{Pdef}) to this system, it is straightforward to show that
\begin{eqnarray}
P(A_{rs}=+1\,|\,r) & = & 0\;,\qquad
P(A_{rs}=-1\,|\,r) \;=\; 1\;,\cr
P(A_{rs}=+1\,|\,s) & = & 1\;,\qquad
P(A_{rs}=-1\,|\,s) \;=\; 0\;,\cr
P(A_{rs}=\pm 1\,|\,t) & = & \frac{1}{2}\;,\qquad\mbox{for $t\neq r, s$}\;,
\end{eqnarray}
and thus,
\begin{eqnarray}
\vev{A_{rs}}_r & = & -1\;,\cr
\vev{A_{rs}}_s & = & +1\;,\cr
\vev{A_{rs}}_t & = & \phantom{-}0\;,\quad\mbox{for $t\neq r, s$.}
\end{eqnarray}
So for each `spin' $A_{rs}$ there exist two `eigenstates':
$\ket{s}$ for $+1$ (`spin' up) and $\ket{r}$ for $-1$ (`spin' down).
For all other states the two outcomes $\pm 1$ are equally probable.

\subsection{Spin Correlations}

A two-`spin' system can be constructed on the tensor product space $\Fq^2\otimes\Fq^2 = \Fq^4$.
The number of non-zero vectors in this space is
$q^4-1$, of which every $q-1$ are equivalent, so the number of inequivalent
states is $(q^4-1)/(q-1)=q^3+q^2+q+1$.
Of these, $(q+1)^2$ are product states, leaving
$(q^3+q^2+q+1)-(q+1)^2=q(q^2-1)$ that are entangled.
Of these, there exists an analog of the spin-singlet state
given by
\begin{equation}
\ket{S} \;=\; \ket{r}\otimes\ket{s}-\ket{s}\otimes\ket{r}\;,\qquad r\neq s\;,
\end{equation}
for any two states $\ket{r}$ and $\ket{s}$ up to a multiplicative constant.
If the characteristic of $\Fq$ is two, the minus sign is replaced by a plus sign.

Products of the `spin' observables are defined as
\begin{equation}
A_{rs}A_{tu}
\,=\,\{
\,\bra{\bar{r}}\otimes\bra{\bar{t}}\,,
\,\bra{\bar{r}}\otimes\bra{\bar{u}}\,,
\,\bra{\bar{s}}\otimes\bra{\bar{t}}\,,
\,\bra{\bar{s}}\otimes\bra{\bar{u}}\,
\}\;,
\label{ArsAtu}
\end{equation}
the four tensor products representing the outcomes
$++$, $+-$, $-+$, and $--$,
and the expectation value giving the correlation between the two `spins.' 
The probabilities of the four outcomes are
particularly easy to calculate for the 
state $\ket{S}$ since \cite{MQT}
\begin{eqnarray}
\bigl(\bra{\bar{r}}\otimes\bra{\bar{s}}\,\bigr)\ket{S}
& =    & \0\quad \mbox{if $r=s$}\;, \cr
& \neq & \0\quad \mbox{if $r\neq s$}\;,
\end{eqnarray}
thus
\begin{equation}
\Bigl|
\bigl(\bra{\bar{r}}\otimes\bra{\bar{s}}\,\bigr)\ket{S}
\Bigr|
\;=\; 1-\delta_{rs}\;,
\end{equation}
and we obtain the probabilities and correlations listed in Table.~\ref{Probs}.
It is straightforward to show that these probabilities cannot be
reproduced by any classical hidden variable theory.\cite{Chang:2012eh,Chang:2012gg}
Thus, GQM is `quantum' in this sense.

\begin{table}[t]
\begin{center}
\begin{tabular}{|c||c|c|c|c||c|}
\hline
\ Observable\ \ &\ $++$\ \ &\ $+-$\ \ &\ $-+$\ \ &\ $--$\ \ &\ E.V. \ \\
\hline
$A_{rs}A_{rs}$ & $0$            & $\dfrac{1}{2}$ & $\dfrac{1}{2}$ & $0$            & $-1$ \phantom{\bigg|} \\
\hline
$A_{rs}A_{rt}$ & $0$            & $\dfrac{1}{3}$ & $\dfrac{1}{3}$ & $\dfrac{1}{3}$ & $-\dfrac{1}{3}$ \phantom{\bigg|}\\
\hline
$A_{rs}A_{st}$ & $\dfrac{1}{3}$ & $\dfrac{1}{3}$ & $0$            & $\dfrac{1}{3}$ & $+\dfrac{1}{3}$ \phantom{\bigg|}\\
\hline
$A_{rs}A_{tu}$ & $\dfrac{1}{4}$ & $\dfrac{1}{4}$ & $\dfrac{1}{4}$ & $\dfrac{1}{4}$ & $\phantom{-}0$  \phantom{\bigg|}\\
\hline
\end{tabular}
\caption{Probabilities and expectation values of product observables in the state $\ket{S}$.
The indices $r$, $s$, $t$, and $u$ are distinct.
Cases that can be obtained by flipping signs using $A_{rs}=-A_{sr}$ are not shown.}
\label{Probs}
\end{center}
\end{table}

It should be noted, though, that GQM also has a common feature with CM when we look at
the Clauser-Horne-Shimony-Holt (CHSH) version of Bell's inequality.\cite{bell,bell2}
The CHSH bound\cite{Clauser:1969ny} is
the upper bound of the absolute value of the following combination of correlators:
\begin{equation}
\vev{A,a\,;B,b}
\;\equiv\;\vev{AB}+\vev{Ab}+\vev{aB}-\vev{ab}
\;,
\label{CHSHcorr}
\end{equation}
where $A$ and $a$ are two observables of particle 1, and
$B$ and $b$ are two observables of particle 2.
All four observables are assumed to take on only the values $\pm 1$ upon 
measurement.
For classical hidden variable theory the bound
on $\left|\vev{A,a;B,b}\right|$ is 2,\cite{Clauser:1969ny} while for
canonical QM it is $2\sqrt{2}$.\cite{cirelson,landau} \footnote{%
The largest possible value of the CHSH bound is 4.\cite{super}
See Ref.~\refcite{Chang:2012we} for a model which saturates this bound.
}
Despite GQM not allowing a hidden variable minic, 
we can nevertheless show, again using Table~\ref{Probs}, that its CHSH bound is the `classical' value of 2.
This bound is independent of the value of $q$ chosen for our Galois field $\Fq$
and is thus also independent of the size of the vector space $\Fq^2\otimes\Fq^2=\Fq^4$.\cite{Chang:2012eh,Chang:2012gg} 
So the limitation on the correlations is not due to the limited number of `spin' direction available
in the model.
This GQM example shows that the absence of hidden variable mimics does not 
necessarily guarantee the violation of the classical CHSH bound.

\subsection{The $q=1$ Limit}

The Galois fields $\Fq$ are only defined for $q=p^n$, that is, 
the order $q$ must be a power of a prime $p$. 
Thus, setting $q=1$ is illegitimate from the Galois field point of view.
Indeed, $\Fq$ consists of $q$ elements so taking the naive $q=1$ limit,
one expects $\Fone$ to consist of only one element (which could be denoted by either $\0$ or $\1$) with no distinction between addition and multiplication.
Such an object is obviously not a ``field.'' 
In other words, a ``field with one element,'' 
in the usual sense of the term, does not and cannot exist. 
However, a different and more interesting picture emerges if 
instead of trying to define $\Fone$ first, 
we set $q=1$ directly in the $N=2$ `spin' model we 
constructed above.

First,
the number of states given in Eq.~(\ref{kets}) will be reduced to $q+1=2$,
\begin{equation}
\ket{\uparrow\;} \;\equiv\; \ket{0}\;=\;\left[\begin{array}{c} \1 \\ \0 \end{array}\right]\;,
\qquad
\ket{\downarrow\;} \;\equiv\; \ket{1}\;=\;\left[\begin{array}{c} \0 \\ \1 \end{array}\right]\;,
\end{equation}
as are the possible outcomes listed in Eq.~(\ref{bras}):
\begin{equation}
\bra{\;\downarrow}\;\equiv\;\bra{\overline{0}}\;=\;\left[\, \0 \;\; \1 \,\right]\;,
\qquad
\bra{\;\uparrow}\;\equiv\;\bra{\overline{1}}\;=\;\left[\, \1 \;\; \0 \,\right]\;.
\end{equation}
%
The sole observable of the system, Eq.~(\ref{Ars}), will be
\begin{equation}
A
\;\equiv\; A_{10} 
\;=\; -A_{01} 
\;=\; \{\;\bra{\overline{1}}\,,\;\bra{\overline{0}}\;\}
\;=\; \{\;\bra{\;\uparrow}\,,\;\bra{\;\downarrow}\;\}
\;,
\end{equation}
for which $\ket{\uparrow\;}$ and $\ket{\downarrow\;}$ are
`eigenstates':
\begin{equation}
\vev{A}_{\uparrow} \;=\; 1\;,\qquad
\vev{A}_{\downarrow} \;=\; -1\;.
\end{equation}
Thus, a measurement of $A$ on $\ket{\uparrow\;}$ will always yield $+1$,
while that on $\ket{\downarrow\;}$ will always yield $-1$.
No superpositions of these states exist, so the system reduces to
a `classical' one where each states has a definite outcome upon measurement.

The `two'-spin system will reduce to
$q^3+q^2+q+1=4$ states, of which $q(q^2-1)=0$ are entangled.
These four are the product states
\begin{equation}
\begin{array}{lll}
\ket{\uparrow\uparrow\;} & \equiv\;
\ket{\uparrow\;}\otimes\ket{\uparrow\;}
\;=\;\left[\begin{array}{c} \1 \\ \0 \\ \0 \\ \0 \end{array}\right]
\;,
\quad
\ket{\uparrow\downarrow\;} & \equiv\;
\ket{\uparrow\;}\otimes\ket{\downarrow\;}
\;=\;\left[\begin{array}{c} \0 \\ \1 \\ \0 \\ \0 \end{array}\right]
\;,\vphantom{\Big|}
\\
\ket{\downarrow\uparrow\;} & \equiv\;
\ket{\downarrow\;}\otimes\ket{\uparrow\;}
\;=\;\left[\begin{array}{c} \0 \\ \0 \\ \1 \\ \0 \end{array}\right]
\;,
\quad
\ket{\downarrow\downarrow\;} & \equiv\;
\ket{\downarrow\;}\otimes\ket{\downarrow\;}
\;=\;\left[\begin{array}{c} \0 \\ \0 \\ \0 \\ \1 \end{array}\right]
\;.\vphantom{\Big|}
\end{array}
\end{equation}
The sole product observable, Eq.~(\ref{ArsAtu}), is
\begin{equation}
AA
\,=\,\{
\,\bra{\;\uparrow\uparrow}\,,
\,\bra{\;\uparrow\downarrow}\,,
\,\bra{\;\downarrow\uparrow}\,,
\,\bra{\;\downarrow\downarrow}\,
\}\;,
\label{AA}
\end{equation}
where
\begin{eqnarray}
\bra{\;\uparrow\uparrow}
& = & \bra{\;\uparrow}\otimes\bra{\;\uparrow}
\;=\; \left[\,\1\;\;\0\;\;\0\;\;\0\,\right]
\,,\cr
\bra{\;\uparrow\downarrow}
& = &
\bra{\;\uparrow}\otimes\bra{\;\downarrow}
\;=\; \left[\,\0\;\;\1\;\;\0\;\;\0\,\right]
\,,\cr
\bra{\;\downarrow\uparrow}
& = &
\bra{\;\downarrow}\otimes\bra{\;\uparrow}
\;=\; \left[\,\0\;\;\0\;\;\1\;\;\0\,\right]
\,,\cr
\bra{\;\downarrow\downarrow}
& = &
\bra{\;\downarrow}\otimes\bra{\;\downarrow}
\;=\; \left[\,\0\;\;\0\;\;\0\;\;\1\,\right]
\,.
\end{eqnarray}
The four states are all `eigenstates' of $AA$ with definite outcomes
when $AA$ is measured.
Thus, the system is `classical' and its CHSH bound is trivially 2.

This discussion can be easily generalized to generic $N$-level GQM,
in which the $q=1$ case reduces to a situation where all surviving states are simultaneous `eigenstates' of all surviving observables.\footnote{%
Note that the vector space $\Fq^N$ has $q^N-1$ non-zero vectors, of which
every $q-1$ of them are physically equivalent in GQM.
Thus, the number of physically distinct states is
$(q^N-1)/(q-1)=[N]_q$,
which reduces to $N$ in the $q=1$ limit.
}
What this exercise shows is that the $q=1$ limit of GQM,
though admittedly fairly trivial, does make sense as a `classical' theory.
Note, however, that in letting $q=1$
we have retained the formalism of GQM as is from the `quantum' $q\neq 1$ cases.
The only thing that has changed is the value of $q$, which
changes the number field over which the state space is constructed,
formally, from $\Fq$ to $\Fone$.
But what is $\Fone$?

\section{The Field with One Element $\bm{\Fone}$}

\subsection{The Observation of Tits}

The concept of $\Fone$ first appeared in a 1957 paper by 
Tits\cite{Tits:1957}
as the ``corps de caract\'eristique 1.''
The main objective of the Tits paper was to use 
projective geometries to define 
the then recently discovered Chevalley groups geometrically.\footnote{%
Chevalley originally used purely algebraic methods in his definition.\cite{Chevalley:1955}}
To each Dynkin diagram and choice of finite field $\Fq$
a projective geometry was associated, and the Chevalley group
identified with the group of projective transformations on the geometry.
For instance, the projective geometry $PG(N-1,q)$ is 
associated with the Dynkin diagram for $A_{N-1}$, and the corresponding
Chevalley group is the projective linear group $PGL(N,q)$.
Towards the end of the paper, Tits argues that the $q=1$ limit of 
$PG(N-1,q)$ actually makes sense as a projective geometry,\footnote{%
With a minor modification to the axioms.  See Ref.~\refcite{Cohn:2004}.}
and that the corresponding group of projective transformations is $S_{N}$.

To see this, let us first list a few properties of
$PG(N-1,q)$ for $q\neq 1$:

\begin{itemize}
\item
The number of points in the geometry $PG(N-1,q)$ is
\begin{equation}
\left[ \begin{array}{c} N \\ 1 \end{array} \right]_q
\;=\; [N]_q
\;=\; \dfrac{q^{N}-1}{q-1} 
\;=\; 1 + q + q^2 + \cdots + q^{N-1}
\;.
\label{Nqdef}
\end{equation}
(See chapter 9 of Ref.~\refcite{Cameron:1994}.)
Here, the notation is:
\begin{equation}
\left[ \begin{array}{c} N \\ M \end{array} \right]_q
\;=\; \dfrac{[N]_q!}{[M]_q! [N-M]_q!}
\;,
\label{Gbinom}
\end{equation}
with 
\begin{eqnarray}
[N]_q  & \equiv & 1 + q + q^2 + \cdots + q^{N-1} \;=\; \dfrac{q^N-1}{q-1}\;,\cr
[N]_q! & \equiv & [N]_q [N-1]_q \cdots [2]_q [1]_q\;.
\label{qAnalogue}
\end{eqnarray}
Eq.~(\ref{Gbinom}) is known as the Gaussian binomial coefficient,
$[N]_q$ the $q$-analogue of the natural number $N$, and $[N]_q!$ the $q$-factorial.

\bigskip

\item
The number of $k$-dimensional subspaces (points, lines, planes, etc.) in $PG(N-1,q)$ are
\begin{eqnarray}
\left[\begin{array}{c} N \\ k+1 \end{array}\right]_q
& = & \dfrac{[N]_q [N-1]_q \cdots [N-k]_q}
            {[k+1]_q [k]_q \cdots [1]_q}
\cr
& = & \dfrac{(q^{N}-1)(q^{N-1}-1)\cdots (q^{N-k}-1)}{(q^{k+1}-1)(q^k-1)\cdots(q-1)}
\;.
\end{eqnarray}
Each subspace contains 
\begin{equation}
[k+1]_q \;=\; \dfrac{q^{k+1}-1}{q-1} \;=\; 1 + q + q^2 + \cdots + q^{k}
\end{equation}
points.

\end{itemize}

\bigskip
\noindent
In the limit $q\rightarrow 1$, the $q$-analog $[N]_q$ reduces to $N$,
the $q$-factorial to the usual factorial, and the Gaussian binomial
coefficients to the usual binomial coefficients.
So the `projective' space $PG(N-1,1)$ should be a space consisting of
$N$ `points,' 
and the number of $k$-dimensional subspaces should
be ${}_NC_{k+1}$, each consisting of $k+1$ points.
This would simply be a set consisting of $N$ elements, and all
the subsets consisting of $k+1$ elements are the $k$-dimensional subspaces.
The group of all possible transformations, projective or not, 
for a set with $N$ elements is obviously $S_N$.

For instance, say $N=3$: $PG(2,1)$ would consist of three `points.'
Let us call them $a$, $b$, and $c$.  The `lines' in this geometry would be
$\{a,b\}$, $\{b,c\}$, and $\{c,a\}$, and the `plane' would be the entire space
$\{a,b,c\}$.
There is a nice discussion in Ref.~\refcite{Cohn:2004}
on how this definition of `geometry' satisfies all the required
axioms.
All possible ways to map the three points onto themselves form the group $S_3$.

Now, recalling that $PG(N-1,q)$ was obtained by identifying the
lines through the origin of $\mathbb{F}_q^{N}$ as points, Tits
argues that the `projective geometry' $PG(N-1,1)$ should be
interpretable as 
\begin{equation}
PG(N-1,1) \;=\; 
\bigl(\,\Fone^{N}\backslash \{\mathbf{\0}\}\,\bigr) \Big/ 
\bigl(\,\Fone \backslash \{\0\}\,\bigr)\;,
\end{equation}
where $\Fone$ is the ``corps de caract\'eristique 1,''
aka the field with one element.

\subsection{$\Fone$ according to Kurokawa and Koyama}

Taken literally, ``corps de caract\'eristique 1'' implies\footnote{%
See footnote \ref{characteristic}.
}
\begin{equation}
\1\;=\;\0\;,
\end{equation}
that is, the multiplicative unit and the additive unit are the same number,
so it may seem that $\Fone=\{\1\}=\{\0\}$, where
\begin{equation}
\1\times\1=\1\;,\qquad \1+\1 \;=\; \1\;.
\label{oneoneone}
\end{equation} 
Such an object would not have any structure to speak of 
and it is difficult to envision how one could construct any geometry,
projective or otherwise, on it. 
Thus, a more sophisticated definition of $\Fone$ is called for.

Various definitions of $\Fone$ and $\Fone$-geometries exist in the literature,
which endeavor to make sense of these concepts,
a survey of which can be found in Ref.~\refcite{LopezPena:2009} by Lorscheid and Lopez-Pena.
Here, we adopt the definition of $\Fone$ by
Kurokawa and Koyama in Ref.~\refcite{Kurokawa:2010} who
argue that $\Fone$ should be interpreted as the set
$\Fone = \{\1\} + \{\0\}$ on which only multiplication is defined.
Note that $\1$ is the multiplicative unit, and 
$\0$ is the multiplicative zero, that is:
\begin{equation}
\1\,\uk \,=\, \uk\,\1 \,=\, \uk\,,\qquad
\0\,\uk \,=\, \uk\,\0 \,=\, \0 \,,\qquad
\uk\,=\,\mbox{$\0$ or $\1$}\;.
\end{equation}
This is different from $\mathbb{F}_2=\{\0,\1\}$ in that
addition is not allowed.\footnote{%
It may be more precise to say that only the addition of $\0$ is allowed.
$\1+\1$ is prohibited.
Another definition referred to in the literature\cite{Zhu:2000,Lescot:2009,Lescot:2011,Lescot:2012}
defines $\1+\1=\1$ with all other 
additions and multiplications among $\0$ and $\1$ defined in the usual way, cf. Eq.~(\ref{oneoneone}).
}
This is the reason for the unfamiliar notation $\{\1\} + \{\0\}$:
$\{\1\}$ is the multiplicative group by itself, and $\{\0\}$ is the multiplicative zero
(but not the additive unit since there is no addition) which is tacked on.
Thus, $\Fone$ is not really a field but what is known as a `monoid.'

Since addition is not allowed, `vector' spaces over $\Fone^{N}$ are not truly `vector' spaces
in the usual sense of the term.
However, lacking a better alternative, let us use vector space terminology anyway.
Since $\Fone^{N}$ is an $N$-dimensional space, 
it will have $N$ basis vectors, which we denote 
\begin{equation}
\ket{1},\;\ket{2},\;\cdots,\;\ket{N}\;.
\end{equation}
The linear combinations of these basis vectors are of the form
\begin{equation}
\sum_{i=1}^{N} c_{i} \ket{i}\;,\qquad c_i\in\Fone\;,
\end{equation}
but since we should have no addition, only one of the $c_i$'s is allowed to be
non-zero at a time.
So the $N$ basis vectors are the only (non-zero) vectors in this space.
They will each be represented by column vectors with only one $\1$, the remaining elements being all $\0$.
Note that these are precisely the states we obtained in taking the limit $q=1$ in GQM.

Linear transformations on $\Fone^N$ must not lead to addition of the basis vectors.
So the only allowed transformations are those for which the $N\times N$ matrix representation has at most
one $\1$ in each row.
The non-singular ones, i.e. the automorphisms, are the ones in which there is one $\1$ in each
row and each column.  These are simply the matrices that permute the $N$ basis vectors.
Since $\Fone\backslash\{\0\}=\{\1\}$, the projective space $PG(N-1,1)$ will simply consist of the
non-zero vectors in $\Fone^{N}$.  
There are $N$ of these, as required for $PG(N-1,1)$.
The projective linear group $PGL(N,1)$ would be $S_{N}$.

Linear maps from $\Fone^N$ to $\Fone$, i.e. the dual vectors, must also satisfy the
no-addition requirement, so they would be $N\times 1$ row-vectors, each with only one $\1$,
the remaining elements all $\0$.
These are precisely the dual vectors representing outcomes in the $q=1$ limit of GQM.

Thus, the Kurokawa-Koyama approach matches well with the GQM construction, and the
$q=1$ `classical' limit can be constructed directly on $\Fone^N$ defined in this fashion.
The resulting state space has the projective geometry $PG(N-1,1)$.
Note, particularly, that the prohibition of addition in Kurokawa-Koyama leads to the
lack of superposition of states in the `classical' limit of GQM.
Therefore, the $q=1$ limit of GQM can be considered to be `classical' due
to this very special feature of $\Fone$.

\section{Summary and Discussion}

Illuminating the precise relation between CM and QM 
remains one of the outstanding problems in the foundations of physics.
In this note we point out that a simple discrete toy model sheds new
light on this important problem. 

In order to compare the two seemingly
disparate formalisms of physics, first, one needs a unifying mathematical
language for such a comparison. This idea, that classical theory should
be rewritten using the mathematics of QM goes back at least to
Koopman\cite{Koopman:1931} and von Neumann.\cite{vNeumann1:1932,vNeumann2:1932} 
In our proposal this unifying mathematical language
is provided by the projective spaces of GQM.\cite{Chang:2012eh,Chang:2012gg}

We have demonstrated that GQM defined on $\Fq^N$ becomes `classical' in the limit $q=1$.
The field with one element $\Fone$, over which this limit can be constructed, 
have all the `classical' properties of consisting only of zero and one, and addition being prohibited.
In other words, the `classical'-ness of $q=1$ GQM can be traced to the
`classical'-ness of $\Fone$.
This way of achieving `classicality' from `quantum'-ness
is quite different from the
usual methodology of taking $\hbar\rightarrow 0$, in the limit of which
operators become commuting.
In GQM, there is no $\hbar$ and there are no operators.

Our result suggests that the pathway from a `quantum' theory to
a `classical' one may not be unique.   
Looking at canonical QM, taking the $\hbar\rightarrow 0$ limit (in some
yet unknown pathway) may not be
the only way for the theory to become classical.
Perhaps replacing $\mathbb{C}$ with some other
mathematical structure, 
perhaps a continuous analog of $\Fone$, though there is no telling such an entity exists,
could lead to a theory which is `classical' in some new way.

The basic premise of this paper is that the relation between CM and QM lies in the differences in the structure and geometry of the underlying number fields upon which the descriptions are predicated.   While we have not discussed the limit $q \to 1$ within the context of the Biorthogonal Quantum Mechanics we defined in Ref.~\refcite{Chang:2012we}, the premise is quite apparent there as well. The approach there was an alternative method of constructing a quantum-like theory on $\Fq^N$.   The CHSH bound on `spin' systems defined on $\Fq^4 = \Fq^2 \times \Fq^2$ in this approach is $q$-dependent.  We were able to demonstrate that this bound takes on the classical value of 2 for $q=3$, and the super-quantum value of 4 for $q=9$.  Once again, it is the differences in the structure of the underlying number fields $\mathbb{F}_3$ and $\mathbb{F}_9=\mathbb{F}_3[\underline{i}]$, 
the latter of which accommodates the analog of the imaginary unit $i$, and their associated geometries, that presage classical, quantum, and super-quantum correlations.\footnote{%
See also Refs.~\refcite{Hanson:2011,Hanson:2012,Hanson:2013}.}

QM is our current fundamental framework of physics.
As such it has passed all available observational tests. And yet
our understanding of QM remains limited. In particular, 
one would still like to understand if QM follows from
few elementary axioms, in analogy what happens in the special theory of relativity. 
Also, one would like to understand in more
detail the precise relation between QM and its classical limit,
the problem addressed in this paper.
Finally, could QM be part of a larger framework, the way
CM appears as a formal limit of QM (or in another
context, similar to the way the special theory of relativity can be naturally
generalized to the general theory of relativity)?
It is very difficult to answer all three questions within the framework of 
canonical QM of the real world as we understand it today, so
a simpler model was called for.
This was the main motivation of our effort to build a toy model of QM using Galois fields.
Surprisingly, all three questions can be addressed in this context. In our previous 
publications on this subject\cite{Chang:2012eh,Chang:2012gg}
we have constructed the precise mathematical rules of the game in this toy world, and we have shed light on the nature of quantum correlations, on the nature of super-quantum correlations\cite{Chang:2012we} (which generalize the usual quantum framework) as well
as on the difference between classical and quantum correlations.
In this paper, we complete this picture by addressing the subtle question of the `classical' limit in the context of GQM.
Understandably enough, our toy world is not the real world. However, for the first time, the two known theories of the real world (the classical and
quantum theory) as well as the conjectured third theory (based on
stronger than quantum, or superquantum, correlations) can be addressed in one coherent theoretical framework, that is simple yet illuminating.
And this is where we find the most useful lessons of our work.

\section*{Acknowledgments}

We would like to thank Chia Hsiung Tze and Cosmas Zachos for helpful discussions.  
DM was supported in part by
the U.S. Department of Energy, grant DE-FG02-13ER41917, task A.
TT is grateful for the hospitality of the Kavli-IPMU during his sabbatical year from fall 2012 to summer 2013
where the initial staged of this work was completed,
and where TT was supported by the World Premier International Research Center Initiative (WPI Initiative), MEXT, Japan.



\begin{thebibliography}{99}


\bibitem{Dirac:1930}
P.~A.~M.~Dirac,
\textit{The Principles of Quantum Mechanics},
(Oxford University Press, 1st edition 1930, 
currently available edition 1982).


\bibitem{Berry:2001}
M.~V.~Berry, 
``Chaos and the semiclassical limit of Quantum mechanics (is the moon there when somebody looks?),''
in \textit{Quantum Mechanics: Scientific perspectives on divine action}, 
editors: R.~J.~Russell, P.~Clayton, K.~Wegter-McNelly and J.~Polkinghorne,
CTNS-Vatican Observatory book series (2001) pp 41-54,
\texttt{http://www.phy.bris.ac.uk/people/berry\_mv/the\_papers/berry337.pdf}.



%
%

\bibitem{Wentzel:1926}
G.~Wentzel, 
``Eine Verallgemeinerung der Quantenbedingungen f�r die Zwecke der Wellenmechanik,''
Zeitschrift f\"ur Physik \textbf{38}, 518--529 (1926).

\bibitem{Kramers:1926}
H.~A.~Kramers,
``Wellenmechanik und halbz\"ahlige Quantisierung,'' 
Zeitschrift f\"ur Physik \textbf{39}, 828--840 (1926).

\bibitem{Brillouin:1926}
L.~Brillouin,
``La m\'ecanique ondulatoire de Schr\"odinger: une m\'ethode g\'en\'erale de resolution par approximations successives,''
Comptes Rendus de l'Academie des Sciences \textbf{183}, 24--26 (1926).


\bibitem{VanVleck:1928}
J.~H.~Van~Vleck,
``The Correspondence Principle in the Statistical Interpretation of Quantum Mechanics,''
Proceedings of the National Academy of Sciences of the USA \textbf{14}, 178--188 (1928).

\bibitem{Schiller:1962a}
R.~Schiller,
``Quasi-Classical Theory of the Nonspinning Electron,''
Phys.\ Rev.\ {\bf 125}, 1100--1108 (1962).

\bibitem{Schiller:1962b}
R.~Schiller,
``Quasi-Classical Transformation Theory,''
Phys.\ Rev.\ {\bf 125}, 1109--1115 (1962).

\bibitem{Schiller:1962c}
R.~Schiller,
``Quasi-Classical Theory of the Spinning Electron,''
Phys.\ Rev.\ {\bf 125}, 1116--1123 (1962).



\bibitem{Bacciagaluppi:2013}
G.~Bacciagaluppi and A.~Valentini,
``Quantum Theory at the Crossroads:
Reconsidering the 1927 Solvay Conference''
(Cambridge University Press, 2013)
[arXiv:quant-ph/0609184]



\bibitem{Bohm:1952a}
D.~Bohm,
``A Suggested Interpretation of the Quantum Theory in Terms of ``Hidden'' Variables I,''
Phys.\ Rev.\ {\bf 85}, 166--179 (1952).

\bibitem{Bohm:1952b}
D.~Bohm,
``A Suggested Interpretation of the Quantum Theory in Terms of ``Hidden'' Variables II,''
Phys.\ Rev.\ {\bf 85}, 180--193 (1952).



\bibitem{Rosen:1964}
N.~Rosen,
``The Relation Between Classical and Quantum Mechanics,''
Am.\ J.\ Phys.\ {\bf 32} (1964) 597--600.

\bibitem{Rosen:1965}
N.~Rosen,
``Mixed States in Classical Mechanics,''
Am.\ J.\ Phys.\ {\bf 33} (1964) 146--150.

\bibitem{Rosen:1986}
N.~Rosen,
``Quantum Particles and Classical Particles,''
Foundations of Physics {\bf 16} (1986) 687--700.



%
%
\bibitem{Koopman:1931}
B.~O.~Koopman,
``Hamiltonian Systems and Transformations in Hilbert Space,''
Proceedings of the National Academy of Sciences of the USA,
\textbf{17} (1931) 315--318.

\bibitem{vNeumann1:1932}
J.~von~Neumann, 
``Zur Operatorenmethode In Der Klassischen Mechanik,''
Annals~of~Mathematics \textbf{33} (3) (1932) 587--642.

\bibitem{vNeumann2:1932}
J.~von~Neumann, 
``Zus\"atze Zur Arbeit `Zur Operatorenmethode...','' 
Annals~of~Mathematics \textbf{33} (4) (1932) 789--791. 

\bibitem{Sudarshan:1976}
 E.~C.~G.~Sudarshan,
 ``Interaction between classical and quantum systems and the measurement of quantum observables,''
 Pramana\ {\bf 6} (1976) 117--126.




%
%

\bibitem{Gozzi:2001he} 
  E.~Gozzi and D.~Mauro,
  ``Minimal coupling in Koopman-von Neumann theory,''
  Annals Phys.\  {\bf 296}, 152--186 (2002)
  [quant-ph/0105113].

\bibitem{Mauro:2001rm} 
  D.~Mauro,
  ``On Koopman-von Neumann waves,''
  Int.\ J.\ Mod.\ Phys.\ A {\bf 17}, 1301--1325 (2002)
  [quant-ph/0105112].

\bibitem{Mauro:2003}
  D.~Mauro,
  ``Topics in Koopman-von Neumann Theory,''
  arXiv:quant-ph/0301172.

\bibitem{Gozzi:2003sh} 
  E.~Gozzi and D.~Mauro,
  ``On Koopman-von Neumann waves 2,''
  Int.\ J.\ Mod.\ Phys.\ A {\bf 19}, 1475--1494 (2004)
  [quant-ph/0306029].

\bibitem{Carta:2005fq} 
  P.~Carta, E.~Gozzi and D.~Mauro,
  ``Koopman-von Neumann formulation of classical Yang-Mills theories. I.,''
  Annalen Phys.\  {\bf 15}, 177--215 (2006)
  [hep-th/0508244].



%
%

\bibitem{Zurek:2003}
W.~H.~Zurek, 
``Decoherence and the Transition from Quantum to Classical -- Revisited,''
in
\textit{Quantum Decoherence, Poincar\'e Seminar 2005},
editors: B.~Duplantier, J.-M. Raimond, and V.~Revasseau, 
Progress in Mathematical Physics Volume~\textbf{48} (Springer, 2007) 1--31
[arXiv:quant-ph/0306072].

\bibitem{Zurek:2003zz} 
  W.~H.~Zurek,
  ``Decoherence, einselection, and the quantum origins of the classical,''
  Rev.\ Mod.\ Phys.\  {\bf 75}, 715 (2003).

\bibitem{Everett:1957hd} 
  H.~Everett,
  ``Relative state formulation of quantum mechanics,''
  Rev.\ Mod.\ Phys.\  {\bf 29}, 454 (1957).
  

%
%

\bibitem{Sherry:1978ea} 
  T.~N.~Sherry and E.~C.~G.~Sudarshan,
  ``Interaction Between Classical and Quantum Systems: A New Approach to Quantum Measurement. 1.,''
  Phys.\ Rev.\ D {\bf 18}, 4580 (1978).

\bibitem{Sherry:1979vc} 
  T.~N.~Sherry and E.~C.~G.~Sudarshan,
  ``Interaction Between Classical And Quantum Systems: A New Approach To Quantum Measurement. Ii. Theoretical Considerations,''
  Phys.\ Rev.\ D {\bf 20}, 857 (1979).
  
\bibitem{Gautam:1978ws} 
  S.~R.~Gautam, T.~N.~Sherry and E.~C.~G.~Sudarshan,
  ``Interaction Between Classical and Quantum Systems: A New Approach to Quantum Measurement. 3. Illustration,''
  Phys.\ Rev.\ D {\bf 20}, 3081 (1979).
  



\bibitem{Peres:2001}
  A.~Peres and D.~R.~Terno,
  ``Hybrid classical-quantum dynamics,''
  Phys.\ Rev.\ A {\bf 63}, 022101 (2001).
      
\bibitem{Terno:2004ti} 
  D.~R.~Terno,
  ``Inconsistency of quantum classical dynamics and what it implies,''
  Found.\ Phys.\  {\bf 36}, 102 (2006)
  [quant-ph/0402092].

\bibitem{Alonso:2010}
  J.~L.~Alonso, A.~Castro, J.~Clemente-Gallardo, J.~C.~Cuch/'i, P.~Echenique, and F.~Falceto,
  ``Geometry and Statistics of Ehrenfest dynamics,''
  arXiv:1010.1494 [quant-ph].

\bibitem{Elze:2011hi} 
  H.~-T.~Elze,
  ``Linear dynamics of quantum-classical hybrids,''
  Phys.\ Rev.\ A {\bf 85}, 052109 (2012)
  [arXiv:1111.2276 [quant-ph]].

\bibitem{Elze:2012fk} 
  H.~-T.~Elze,
  ``Four questions for quantum-classical hybrid theory,''
  J.\ Phys.\ Conf.\ Ser.\  {\bf 361}, 012004 (2012)
  [arXiv:1202.3448 [quant-ph]].
  
\bibitem{Elze:2013una} 
  H.~-T.~Elze,
  ``Quantum-classical hybrid dynamics � a summary,''
  J.\ Phys.\ Conf.\ Ser.\  {\bf 442}, 012007 (2013).

\bibitem{Jauslin:2011}
H.~R.~Jauslin and D.~Signy,
``Dynamics of mixed classical-quantum systems, geometric quantization and coherent states,''
in \textit{Mathematical Horizons for Quantum Physics}, 
edited by H.~Araki, B.-G.~Englert, L.-C.~Kwek, and J.~Suzuki,
Lecture Notes Series, Institute for Mathematical Sciences, 
National University of Sinpapore, Vol. 20 (World Scientific, Singapore, 2010)
[arXiv:1111.5774 [quant-ph]].

\bibitem{Salcedo:2012qx} 
  L.~L.~Salcedo,
  ``Statistical consistency of quantum-classical hybrids,''
  Phys.\ Rev.\ A {\bf 85}, 022127 (2012)
  [arXiv:1201.4237 [quant-ph]].

\bibitem{Barcelo:2012ja} 
  C.~Barcelo, R.~Carballo-Rubio, L.~J.~Garay and R.~Gomez-Escalante,
  ``Hybrid classical-quantum formulations ask for hybrid notions,''
  Phys.\ Rev.\ A {\bf 86}, 042120 (2012)
  [arXiv:1206.7036 [quant-ph]].






%
%

\bibitem{Chang:2012eh} 
  L.~N.~Chang, Z.~Lewis, D.~Minic and T.~Takeuchi,
  ``Galois Field Quantum Mechanics,''
  Mod.\ Phys.\ Lett.\ B {\bf 27}, 1350064 (2013)
  [arXiv:1205.4800 [quant-ph]].

\bibitem{Chang:2012gg} 
  L.~N.~Chang, Z.~Lewis, D.~Minic and T.~Takeuchi,
  ``Spin and Rotations in Galois Field Quantum Mechanics,''
  J.\ Phys.\ A:\ Math.\ Theor.\ {\bf 46}, 065304 (2013)
  [arXiv:1206.0064 [quant-ph]].




%
%

\bibitem{MQT}
 B.~Schumacher and M.~D.~Westmoreland,
 ``Modal Quantum Theory,''
 Foundations of Physics {\bf 42} (2012) 910--925
 [arXiv:1010.2929 [quant-ph]].

\bibitem{MQT2}
 B.~Schumacher and M.~D.~Westmoreland,
 ``Non-contextuality and free will in modal quantum theory,''
 arXiv:1010.5452 [quant-ph].
 
\bibitem{MQT3}
 M.~D.~Westmoreland and B.~Schumacher,
 ``Possibility, probability and entanglement: Non-contextuality in modal quantum mechanics,''
 in the Proceedings of \textit{Foundations of Probability and Physics 6} (FPP6), June 13-16, 2011, V\"axj\"o, Sweden, 
 AIP Conference Proceedings 1424, pp. 364-368 (2012).

\bibitem{MQT4}
 B.~Schumacher and M.~D.~Westmoreland,
 ``Almost quantum theory,''
 arXiv:1204.0701 [quant-ph].



%
%

\bibitem{James:2011}
 R.~P.~James, G.~Ortiz, and A. Sabry,
 ``Quantum Computing over Finite Fields,''
 arXiv:1101.3764 [quant-ph].
 
\bibitem{Willcock:2011}
 J.~Willcock and A.~Sabry,
 ``Solving UNIQUE-SAT in a Modal Quantum Theory,''
 arXiv:1102.3587 [quant-ph].
 
\bibitem{Hanson:2011}
 A.~J.~Hanson, G.~Ortiz, A.~Sabry, and J.~Willcock,
 ``The Power of Discrete Quantum Theories,''
 arXiv:1104.1630 [quant-ph].
 
\bibitem{Hanson:2012}
 A.~J.~Hanson, G.~Ortiz, A.~Sabry, and Y.-T. Tai,
 ``Geometry of Discrete Quantum Computing,''
 J.\ Phys.\ A:\ Math.\ Theor.\ {\bf 46}, 185301 (2013)
 [arXiv:1206.5823 [quant-ph]].
 
\bibitem{Hanson:2013}
 A.~J.~Hanson, G.~Ortiz, A.~Sabry, and Y.-T. Tai,
 ``Discrete Quantum Theories,''
 J.\ Phys.\ A:\ Math.\ Theor.\ {\bf 47}, 115305 (2014)
 [arXiv:1305.3292 [quant-ph]].




%
%

\bibitem{Chang:2012we} 
  L.~N.~Chang, Z.~Lewis, D.~Minic and T.~Takeuchi,
  ``Biorthogonal Quantum Mechanics: Super-Quantum Correlations and Expectation Values without Definite Probabilities,''
  J.\ Phys.\ A:\ Math.\ Theor.\ {\bf 46}, 485306 (2013)
  [arXiv:1208.5189 [math-ph]].

\bibitem{Takeuchi:2012mra} 
  T.~Takeuchi, L.~N.~Chang, Z.~Lewis and D.~Minic,
  ``Some Mutant Forms of Quantum Mechanics,''
  in the Proceedings of \textit{Quantum Theory: Reconsideration of Foundations 6} (QTRF6), 
  June 11-14, 2012, V\"axj\"o, Sweden,
  AIP Conference Proceedings 1508
  [arXiv:1208.5544 [quant-ph]].

\bibitem{Chang:2013joa} 
  L.~N.~Chang, Z.~Lewis, D.~Minic and T.~Takeuchi,
  Int.\ J.\ Mod.\ Phys.\ A {\bf 29}, 1430006 (2014),
  and
  in the Proceedings of \textit{the Conference in Honour of the 90th Birthday of Freeman Dyson}, 26-29 August 2013, Institute of Advanced Studies at Nanyang Technological University, Singapore
  [arXiv:1312.0645 [quant-ph]].



%
%

\bibitem{Ellerman:2013}
 D.~Ellerman,
 ``Quantum mechanics over sets,''
 arXiv:1310.8221 [quant-ph].

\bibitem{Ellerman:2014}
 D.~Ellerman,
 ``Partitions and Objective Indefiniteness in Quantum Mechanics,''
 arXiv:1401.2421 [quant-ph].



%
%

\bibitem{Curtright:2005zk} 
  T.~Curtright and L.~Mezincescu,
  ``Biorthogonal quantum systems,''
  J.\ Math.\ Phys.\  {\bf 48}, 092106 (2007)
  [quant-ph/0507015].

\bibitem{Curtright:2006zv} 
  T.~Curtright, L.~Mezincescu and D.~Schuster,
  ``Supersymmetric biorthogonal quantum systems,''
  J.\ Math.\ Phys.\  {\bf 48}, 092108 (2007)
  [quant-ph/0603170].



%
%

\bibitem{Tits:1957}
J.~Tits, 
``Sur les analogues algebriques des groupes semi-simple complexes,''
Colloque d'algebre superieure, tenu a Bruxelles du 19 au 22 decembre 1956,
Centre Belge de Recherches Math., Gauthier-Villar, Paris 1957, 261--289.

\bibitem{qAnalog}
E.~W.~Weisstein, ``$q$-Analog,'' 
from MathWorld -- a Wolfram Web Resource,
\texttt{http://mathworld.wolfram.com/q-Analog.html}.


\bibitem{Manin:1995}
Y.~I.~Manin,
``Lectures on zeta functions and motives (according to Deninger and Kurokawa),''
Ast\'erisque {\bf 228} (1995) 121-163.

\bibitem{Manin:2008}
Y.~I.~Manin,
``Cyclotomy and analytic geometry over $\Fone$,''
arXiv:0809.1564 [math.AG].

\bibitem{Zhu:2000}
Y.~Zhu,
``Combinatorics and characteristic one algebra,'' 
preprint, 2000.

\bibitem{Soule:2003}
C.~Soul\'e,
``Les vari\'et\'es sur le corps \`a un \'el\'ement,''
Moscow\ Math.\ J.\ {\bf 4} (2004) 217-244
[arXiv:math/0304444v1 [math.AG]].

\bibitem{Soule:2009}
C.~Soul\'e,
``Lectures on algebraic varieties over $\Fone,$''
in \textit{Noncommutative Geometry, Arithmetic, and Related Topics,
Proceedings of the Twenty-First Meeting of the Japan-U.S. Mathematics Institute},
edited by C.~Consani and A.~Connes (Johns Hopkins University Press, 2012)
267--278.



\bibitem{Kurokawa:2003}
N.~Kurokawa, H.~Ochiai, and M.~Wakayama,
``Absolute derivations and zeta functions,''
Documenta Math, Extra Volume: Kazuya Kato's Fiftieth Birthday (2003) 
565--584.

\bibitem{Kurokawa:2005}
N.~Kurokawa,
``Zeta functions over $\Fone$,''
Proc.\ Japan\ Acad.\ {\bf 81}, Ser. A (2005) 180-184.

\bibitem{Koyama:2012}
S.~Koyama and N.~Kurokawa,
``Absolute zeta functions and absolute tensor products,''
in \textit{Noncommutative Geometry, Arithmetic, and Related Topics,
Proceedings of the Twenty-First Meeting of the Japan-U.S. Mathematics Institute},
edited by C.~Consani and A.~Connes (Johns Hopkins University Press, 2012)
225--240.


\bibitem{Deitmar:2005}
A.~Deitmar,
``Schemes over $\Fone$,''
in \textit{Number Fields and Function Fields -- Two Parallel Worlds},
edited by G.~van der Geer, B.~J.~J.~Moonen, and R.~Schoof,
Prog.\ Math.\ {\bf 239} (2005) pp. 87-100
[arXiv:math/0404185 [math.NT]].

\bibitem{Deitmar:2006a}
A.~Deitmar,
``Remarks on zeta functions and $K$-theory over $\Fone$,''
Proc.\ Japan\ Acad.\ {\bf 82}, Ser. A (2006) 141--146
[arXiv:math/0605429 [math.NT]].

\bibitem{Deitmar:2006b}
A.~Deitmar,
``$\Fone$-schemes and toric varieties,''
Contributions to Algebra and Geometry {\bf 49} (2008)
pp. 517--525
[arXiv:math/0608179v9 [math.NT]].

\bibitem{Deitmar:2008}
A.~Deitmar, S.~Koyama, and N.~Kurokawa,
``Absolute zeta functions,''
Proc.\ Japan\ Acad.\ {\bf 84}, Ser. A (2008) 138-142.

\bibitem{Borger:2009}
J.~Borger,
``$\Lambda$-rings and the field with one element,''
arXiv:0906.3146v1 [math.NT].

\bibitem{Lorscheid:2009}
O.~Lorscheid,
``Algebraic groups over the field with one element,''
Math.~Zeitschrift {\bf 271} (2012) 117--138 
[arXiv:0907.3824v1 [math.AG]].

\bibitem{LopezPena:2009}
J.~L\'opez~Pe\~na and O.~Lorscheid,
``Mapping $\Fone$-Land: An overview of geometries over the
field with one element,''
in \textit{Noncommutative Geometry, Arithmetic, and Related Topics,
Proceedings of the Twenty-First Meeting of the Japan-U.S. Mathematics Institute},
edited by C.~Consani and A.~Connes (Johns Hopkins University Press, 2012)
241--266
[arXiv:0909.0069v1 [math.AG]].

\bibitem{LeBruyn:2009}
L.~Le~Bruyn,
``(Non)Commutative $\Fun$ Geometry,''
arXiv:0909.2522v1 [math.RA].



\bibitem{Connes:2008a}
A.~Connes, C.~Consani, and M.~Marcolli,
``Fun with $\Fone$,''
arXiv:0806.2401v1 [math.AG].

\bibitem{Connes:2008b}
A.~Connes and C.~Consani,
``On the notion of geometry over $\Fone$,''
J.~Algebraic.~Geom. {\bf 20} (2011) 525-557
[arXiv:0809.2926v2 [math.AG]].

\bibitem{Connes:2009a}
A.~Connes and C.~Consani,
``Schemes over $\Fone$ and Zeta Functions,''
[arXiv:0903.2024v3 [math.AG]].

\bibitem{Connes:2009b}
A.~Connes and C.~Consani,
``Characteristic 1, Entropy and the Absolute Point,''
in \textit{Noncommutative Geometry, Arithmetic, and Related Topics,
Proceedings of the Twenty-First Meeting of the Japan-U.S. Mathematics Institute},
edited by C.~Consani and A.~Connes (Johns Hopkins University Press, 2012)
75--140
[arXiv:0911.3537v1 [math.AG]].

\bibitem{Connes:2010a}
A.~Connes and C.~Consani,
``From Monoids to Hyperstructures: In Search of an Absolute Arithmetic,''
[arXiv:1006.4810v1 [math.AG]].

\bibitem{Connes:2010b}
A.~Connes,
``The Witt Construction in Characteristic One and Quantization,''
[arXiv:1009.1769v1 [math.QA]].

\bibitem{Lescot:2009}
P.~Lescot,
``Alg\`ebre absolue,''
Annales des sciences math\'ematiques du Qu\'ebec {\bf 33}, 1 (2009) 63--82
[arXiv:0911.1989 [math.RA]].

\bibitem{Lescot:2011}
P.~Lescot,
``Absolute Algebra II -- Ideals and spectra''
J. Pure Applied Algebra {\bf 215}, 7 (2011) 1782--1790.

\bibitem{Lescot:2012}
P.~Lescot,
``Absolute Algebra III -- The saturated spectrum''
J. Pure Applied Algebra {\bf 216}, 5 (2012) 1004--1015
[arXiv:1310.8399 [math.RA]].





\bibitem{Kurokawa:2010}
N.~Kurokawa and S.~Koyama,
\textit{Absolute Mathematics}
(Nippon Hy\=oron Sha, 2010) (in Japanese).

\bibitem{Kurokawa:2013}
N.~Kurokawa and S.~Koyama,
\textit{Intoduction to the ABC Conjecture}
(PHP Institute, 2013) (in Japanese).

\bibitem{Mochizuki:2012}
S.~Mochizuki,\\
\texttt{http://www.kurims.kyoto-u.ac.jp/$\sim$motizuki/papers-english.html}.

\bibitem{BerryKeating:1998}
M.~V.~Berry and J.~P.~Keating,  
``$H = xp$ and the Riemann zeros,'' 
in 
\textit{Supersymmetry and Trace Formulae: Chaos and Disorder}, 
Editors: I.~V.~Lerner, J.~P.~Keating and D.~E.~Khmelnitskii,
NATO Science Series B, Physics Vol.~370 (Springer, 1999). 
 
  

\bibitem{Hirschfeld}
 J.~W.~P.~Hirschfeld, {\it Projective Geometries over Finite Fields,} 2nd ed.
 (Oxford University Press, 1998).

\bibitem{Hirschfeld2}
 J.~W.~P.~Hirschfeld, G.~Korchm\'aros, and F.~Torres,
 {\it Algebraic Curves over a Finite Field}
 (Princeton University Press, 2008).

\bibitem{Arnold}
 V.~I.~Arnold, {\it Dynamics, Statistics and Projective Geometry of Galois Fields}
 (Cambridge University Press, 2011).

\bibitem{Ball-Weiner}
 S.~Ball and Z.~Weiner, {\it An Introduction to Finite Geometry,}\\
 \texttt{http://www-ma4.upc.es/\~{}simeon/IFG.pdf}


\bibitem{bell}
 J.~S.~Bell, 
 ``On the Einstein Podolsky Rosen paradox,''
 Physics {\bf 1} (1964) 195.

\bibitem{bell2}
 J.~S.~Bell, {\it Speakable and Unspeakable in Quantum Mechanics}, Cambridge University Press (1987).

\bibitem{Clauser:1969ny}
  J.~F.~Clauser, M.~A.~Horne, A.~Shimony and R.~A.~Holt,
  ``Proposed experiment to test local hidden variable theories,''
  Phys.\ Rev.\ Lett.\  {\bf 23} (1969) 880.

\bibitem{cirelson}
 B. S. Cirel'son, 
 ``Quantum Generalizations of Bell's Inequality,''
 Lett. Math. Phys. {\bf 4} (1980) 93.

\bibitem{landau}
 L. J. Landau, 
 ``On the violation of Bell's inequality in quantum theory,''
 Phys. Lett. {\bf A 120} (1987) 52.

\bibitem{super}
 S.~Popescu and D.~Rohrlich, 
 ``Nonlocality as an Axiom,''
 Foundations of Physics, {\bf 24} (1994) 379.



\bibitem{Chevalley:1955}
C.~Chevalley, 
``Sur certains groups simples,''
T\=ohoku Math. Journ. {\bf 7} (2) (1955), 14--66.

\bibitem{Cohn:2004}
H.~Cohn,
``Projective Geometry over $\Fone$ and the Gaussian Binomial Coefficients,''
American Mathematical Monthly {\bf 111} (2004) 487--495
[arXiv:math/0407093 [math.CO]].

\bibitem{Cameron:1994}
P.~J.~Cameron,
\textit{Combinatorics: Topics, Techniques, Algorithms}
(Cambridge University Press, 1994).

\bibitem{Penrose:1971}
R.~Penrose, 
``Angular momentum: an approach to combinatorial space-time,''\\
in \textit{Quantum Theory and Beyond, Essays and Discussions Arising from a Colloquium},\\ 
edited by T.~Bastin (Cambridge University Press, 1971)\\
\texttt{http://math.ucr.edu/home/baez/penrose/Penrose-AngularMomentum.pdf}


\bibitem{Penrose:1972}
R.~Penrose, 
``On the nature of quantum geometry,''\\
in \textit{Magic Without Magic: John Archbald Wheeler},\\
edited by J.~R.~Klauder (Freeman, 1972) p.~333--354,\\
\texttt{http://math.ucr.edu/home/baez/penrose/Penrose-OnTheNatureOfQuantumGeometry.pdf}


\end{thebibliography}
\end{document}